\documentclass[draft]{agujournal2019}
\usepackage{url} 
\usepackage[inline]{trackchanges} 
\usepackage{soul}
\usepackage{float}

\usepackage{titlesec}
\usepackage{array, booktabs, caption}
\usepackage{makecell}
\usepackage{color, colortbl}
\usepackage{rotating}
\usepackage{gensymb}
\usepackage{float}

\newcolumntype{P}[1]{>{\centering\arraybackslash}p{#1}}
\newcolumntype{M}[1]{>{\centering\arraybackslash}m{#1}}
\usepackage[T1]{fontenc}
\usepackage{babel}
\pagecolor{white}

\defineshorthand{"=}{\penalty1000-\hskip0pt}

\draftfalse
\journalname{JGR: Atmospheres}

\begin{document}

\title{Upper Atmosphere Smoke Injection from Large Areal Fires}

\authors{Stephanie Redfern\affil{1}, Julie K. Lundquist\affil{1,2}, Owen B. Toon\affil{1,3}, Domingo Mu\~{n}oz-Esparza\affil{4},  Charles G. Bardeen \affil{1,4}, Branko Kosovi\'{c}\affil{4}}

\affiliation{1}{Department of Atmospheric and Oceanic Sciences, University of Colorado, Boulder, Colorado, USA}
\affiliation{2}{National Renewable Energy Laboratory, Golden, Colorado, USA}
\affiliation{3}{Laboratory for Atmospheric and Space Physics, Boulder, Colorado, USA}
\affiliation{4}{National Center for Atmospheric Research, Boulder, Colorado, USA}
\correspondingauthor{Stephanie Redfern}{stephanie.redfern@colorado.edu}

\begin{keypoints}
\item	Local atmospheric conditions affect plume lofting above large areal fires simulated by WRF-Fire
\item	Moisture has the largest impact on the amount of smoke lofted into the upper troposphere and lower stratosphere
\item	Faster wind speeds reduce the altitude of smoke lofting
\end{keypoints}


\begin{abstract}
Large fires can inject smoke into the upper troposphere and lower stratosphere. Detailed fire simulations allow for assessment of how local weather interacts with these fires and affects smoke lofting. In this study, we employ the fire simulation package in the Weather Research and Forecasting model (WRF-Fire), Version 4.0.1, to explore how smoke lofting from a fire burning a homogeneous fuel bed changes with varying local winds, relative humidity, and atmospheric boundary-layer stability for two different-sized areal fires. We find that the presence of moisture has the greatest influence on the results by raising the altitude of lofting, while faster winds speeds dampen lofting and lower the injection height. Stably stratified conditions further inhibit plume propagation compared with neutrally stratified conditions.

\end{abstract}


\section{Introduction}

Large fires have the potential to loft considerable amounts of smoke high in the atmosphere. If the fires burn intensely enough, their sensible heat flux can generate powerful convection \cite{luderer_modeling_2006,trentmann_modeling_2006}. Convection can also be enhanced by cloud formation if the atmosphere is humid. Aerosol byproducts of combustion can act as cloud condensation nuclei (CCNs), which can seed pyrocumulus (fire-derived clouds, or pyroCu) and pyrocumulonimbus (fire-derived storms, or pyroCb) development. PyroCb in particular can generate deep convection, which has the potential to inject smoke into the upper atmosphere and stratosphere. When aerosols are lofted, high-speed, upper-level winds can advect them to locations far from the fire source before they are removed from the atmosphere, and therefore they can affect climate elsewhere on a months-long timescale \cite{fromm,peterson_quantifying_2014,peterson,yu}. 

The impact of smoke transport on climate has been a research topic for decades. In the late 20th century, Cold War tensions between the United States and the Soviet Union presented the risk of nuclear war. The 70,000 nuclear weapons then in existence exceeded the numbers of military targets by orders of magnitude, so urban areas were almost certain to be targeted. In an early research study, \citeA{crutzen} suggested that urban attacks might cause fires in surrounding forested regions, leading to very dense but short-lived smoke palls \cite{crutzen}. Another early study, \citeA{turco_global}, found that whole cities\textemdash{}with much larger fuel loads than forests\textemdash{}could be ignited, with powerful enough convection to loft combustion byproducts into the upper troposphere and stratosphere, where residence times could be on the order of years \cite{turco_global}. In a few weeks' time, these aerosols would circle the planet and limit sunlight globally. As a result, following significant drops in temperature and ensuing crop failure, people would either freeze or starve to death. This outcome was given a name: nuclear winter \cite{turco_global}. Despite denuclearization efforts over the last forty years, nine nations still possess a total of 14,000 nuclear weapons, of which about 5000 are deployed (with the rest being dismantled) \cite{SIPRI}. This stockpile is enough to attack with more than 10 weapons each city with over 100,000 people in the U.S., Russia, and China \cite{kristensen_chapter_2019,arms_control_assn}. Nuclear winter, therefore, remains a risk.

More recent research has employed both global climate models (GCMs) and high-resolution models to assess these possible climate impacts, and they have by and large corroborated the findings of early studies \shortcite{covey1984global,cotton,penner,ghan1988climatic,robock_nuclear_2007,robock_climatic_2007,toon_regional,robock_sadestruction,reisner,coupe}. Most GCM-based studies are initialized with a certain amount of smoke loading at a specified injection height, based on the size and number of weapons considered. These altitudes are often informed by observations from mass fires, such as in Hamburg during the second world war, or on 1980's simulations using mesoscale models that were driven by energy release rates \cite{penner,small_early_1988,turco1990climate}. Using this method, \citeA{robock_climatic_2007}, \citeA{toon_regional}, \citeA{toon_environmental_2008}, and \citeA{coupe} found that, given an exchange with the strategic weapons allowed under treaties between the U.S. and Russia, a nuclear winter is a possible outcome. Further, the detonation of 100 Hiroshima-sized bombs in Pakistani and Indian urban areas could produce climate effects that endanger global agricultural productivity \cite{xia2015decadal,toon_rapidly_2019,jagermeyr2020regional}. Only 1.5 megaton's worth of explosives used on urban areas could cool global average temperatures by 1.25{\degree}C for several years and cause anomalously cold temperatures of 0.5{\degree}C for over a decade \cite{toon_rapidly_2019}. In contrast to these findings, \citeA{reisner}, using a low-fuel loading scenario of an estimated 0.14 g cm\textsuperscript{-2} within the target area \cite{robock_comment_2019} and 0.91 g cm\textsuperscript{-2} across the entire domain, suggests that the risk of significant climate perturbations is small because the bulk of the smoke rising from urban targets will never ascend above the lower troposphere. This particular study used a fine-resolution model to simulate a detonation in the East Lake country club region of Atlanta \cite{robock_comment_2019} to predict a smoke concentration profile, which was then applied in a global climate model to more generally assess the impacts of high-altitude smoke transport following a broader-scale nuclear exchange \cite{reisner}. The fuel sources were based on satellite imagery of the region and idealized atmospheric conditions of 6 m s\textsuperscript{-1} surface winds and a stable and dry atmosphere were used in the microscale model. 

Some studies focused on local smoke lofting, rather than on global climatic effects, have indicated that weather plays a significant role in fire plume behavior, which affects how much smoke is injected in the upper atmosphere. PyroCu and pyroCb development is sensitive to local environmental moisture \cite{fromm,lareau_clements_2016,peterson_conceptual_2016}, and as previously mentioned, these phenomena can generate deep convection. Additionally, wind speeds determine whether or not a mass fire develops into a conflagration or a firestorm\textemdash{}the latter of which causes much deeper smoke lofting \cite{glasstone_effects_1977,cotton,penner}. In the context of climatic forcing by upper atmosphere aerosols, it is important, therefore, to have localized estimations of plume rise.

In this study, we use large-eddy simulations (LES) to simulate fire growth in a homogenous, wildland fuel bed to evaluate smoke lofting due to mass fire development, on time scales relevant to plume behavior. We quantify the sensitivity of this lofting to varying atmospheric conditions. Two fire sizes are examined, informed by the presumed affected areas following a small-sized (15 kiloton) nuclear weapon air burst, which is consistent with the Hiroshima nuclear explosion, and a larger weapon (100 kiloton), which could be present in the Indian arsenal, since they exploded a weapon with a yield near 50 kt in 1998 \cite{nukemap}. The makeup of the fuel bed is less dense than what would be found in an urban setting, however. Fuel loading, at 1.3 g cm\textsuperscript{-2}, is an order of magnitude or more lower than what would be expected if the fire area consisted of buildings, refineries, and other similar targets. Consequently, this smoke sensitivity study primarily focuses on how local weather conditions affect plume rise. Longer-term radiatively-driven lofting, which has been seen in observations \citeA{yu} and climate models, is not considered.

The simulations and model setup are described in Section 2. In Section 3, we address the role of oxygen starvation, present our findings on how wind speeds, atmospheric moisture, and boundary-layer stability affect vertical smoke distributions, and compare several metrics to evaluate the comparative sensitivity of plume rise to winds and humidity. Finally, we provide a brief discussion and conclusions in Section 4.


\section{Model Environment}
The simulations use the Advanced Research Weather Research and Forecasting Model (WRF-ARW, henceforth referred to as WRF), Version 4.0.1 \cite{skamarock_2008,powers_weather_2017}. WRF is a numerical weather prediction model capable of simulating nested domains with outer boundaries specified either by coarser-scale models or reanalysis data, or by idealized conditions\textemdash{}the latter of which is useful for parameter sensitivity studies like this one. As an open-source model, WRF offers ease of accessibility and, therefore, simulation reproducibility. The model setup, explained below, is also summarized in Table \ref{modenv}.

WRF-Fire, the fire modeling parameterization in WRF, simulates the fire spread and heat release (latent and sensible) that results from a wildland fire. This heat release feeds back into the atmospheric component of WRF, influencing local meteorology. A passive smoke tracer option may be activated. WRF-Fire was first introduced in 2004 as an integration into WRF that coupled the CAWFE tracer model with a level-set fire spread algorithm \cite{clark,patton,mandel_2011,coen_2013}. In this study, we use a more recent version of WRF-Fire, in which a level-set method, which is a mathematical algorithm used to track front propagation \cite{osher_fronts_1988}, is used to determine the fire spread. This technique improves the accuracy of its fire-front tracking by incorporating high-order numerical schemes and solutions of an additional equation for level-set re-initializations \cite{me_wrfire,jimenez_high_2018}. 


\subsection{Model Setup}

As described in Table \ref{modenv}, the simulations include two domains: an outer mesoscale domain with periodic boundary conditions, acting as the reference flow, and a nested LES. The outer domain has a horizontal grid resolution of 1.1 km across a 200 x 200 km domain. The LES nest is centered inside the parent, with a horizontal grid resolution of 100 m and spanning a 19.8 km x 19.8 km domain. The vertical resolution for both domains is the same, with 80 levels stretched between the ground and the model top at 18 km. Due to the low model top, we assume a mid-latitude location with a tropopause at 12 km and a Coriolis parameter of 10\textsuperscript{-4} s\textsuperscript{-1} \cite{trop_2012}. The fire mesh, a subgrid of the nested domain used with the fire spread algorithm, has a 25-m grid resolution. The fire is lit only in the nest, and feedback from the nest to the parent domain does not occur. The simulations automatically terminate when the fire reaches the edge of the LES domain, which could occur as quickly as 1.5 hours for the high wind speed simulations used in this study.

\begin{table}[ht]
    \centering
    \begin{tabular}{ | M{3cm} | M{5cm} | M{5cm} |} 
       \hline  \rowcolor{white}
       \thead{WRF Parameter} & \thead{d01} & \thead{d02} \\ \hline \rowcolor{lightgray}
       Horizontal Resolution & 1.1 km & 100 m \\ \hline \rowcolor{white}
       Horizontal Extent (Grid Cells) & 200 x 200 & 199 x 199 \\ \hline \rowcolor{lightgray}
       Number of Vertical Levels & \multicolumn{2}{c|}{80}  \\ \hline \rowcolor{white}
       Model Top & \multicolumn{2}{c|}{18 km} \\ \hline  \rowcolor{lightgray}
       Lowest Vertical Level & \multicolumn{2}{c|}{60 m} \\ \hline \rowcolor{white}
       Time Step & 3 s & 0.27 s \\ \hline  \rowcolor{lightgray}
       Time Integration Scheme & \multicolumn{2}{c|}{Runge-Kutta 3rd Order} \\  \hline \rowcolor{white}
       Advection Scheme & \multicolumn{2}{c|}{5th-order Horizontal; 3rd order Vertical} \\  \hline \rowcolor{lightgray}
       Coriolis Parameter &  \multicolumn{2}{c|}{10\textsuperscript{-4} s\textsuperscript{-1}} \\ \hline \rowcolor{white}
       Planetary Boundary Layer Scheme & MYNN & - \\ \hline \rowcolor{lightgray}
       Eddy Coefficient Option & 2D horizontal diffusion & 1.5-order SGS TKE \\ \hline \rowcolor{white}
       Surface Layer Scheme & \multicolumn{2}{c|}{MYNN} \\ \hline \rowcolor{lightgray}
       Turbulence \& Mixing & - & Isotropic with 1.5-order TKE closure scheme \\ \hline \rowcolor{white}
       Damping & \multicolumn{2}{c|}{Upper-level Rayleigh layer of 0.33 s\textsuperscript{-1} with a depth of 5 km} \\ \hline \rowcolor{lightgray}
       Roughness Length & \multicolumn{2}{c|}{0.3 m} \\ \hline \rowcolor{white}
       Microphysics Scheme & \multicolumn{2}{c|}{Thompson} \\ \hline \rowcolor{lightgray}
       Boundary Conditions & Periodic & Interpolated, One-way Nested \\ \hline \rowcolor{white}
       Fire Grid Mesh & - & 4:1 \\ \hline \rowcolor{lightgray}
       Ignition Radius & - & 2-km or 4-km \\ \hline \rowcolor{white}
       Fuel Source & - & Heavy Logging Slash,  0.9 m depth (1.3 g cm\textsuperscript{-1}) \\ \hline \rowcolor{lightgray}
       Smoke Handling & - & Passive Tracer \\ \hline  \rowcolor{white}
    \end{tabular} 
    \caption{WRF Configuration}
    \label{modenv}
\end{table}

WRF physics and dynamics are simplified. The outer domain uses the Mellor-Yamada-Nakanishi-Niino (MYNN) PBL scheme \cite{mynn}. Since the nest is configured with LES resolution, no planetary boundary layer (PBL) scheme is required or specified in this domain. Both domains use Thompson graupel microphysics \cite{thompson,thompson_2008}.  The MYNN surface-layer scheme is used to account for near-surface heating impacts, but no land-surface parameterization is used. For simulations incorporating radiative fluxes, the Rapid Radiation Transfer Model for the Korean Integrated Model (RRTMK) short- and longwave schemes are enabled \cite{baek_revised_2017}. The model is initialized and run in a non-hydrostatic environment. A third-order Runge-Kutta time integration scheme is used, with a time step of 0.5 s in the outer domain. Eddy diffusion is handled by horizontal Smagorinsky first-order closure for the mesoscale domain, and by 1.5 order TKE closure for the LES nest.  Moisture and scalar advection variables are calculated via a positive-definite scheme. Fifth-order horizontal and third-order vertical advection accuracy is used. To reduce the impact of gravity wave reflection within the domain, upper-level Rayleigh damping with a coefficient of 0.2 s\textsuperscript{-1} is employed within a depth of 5 km from the model top \cite{klemp_upper_2008}. This damping layer is not expected to affect smoke lofting within the vertical range of interest in this study.


\subsection{Simulations Overview}

\begin{figure}[h!]
\centering
\hspace*{-1cm}
\includegraphics[width=16cm]{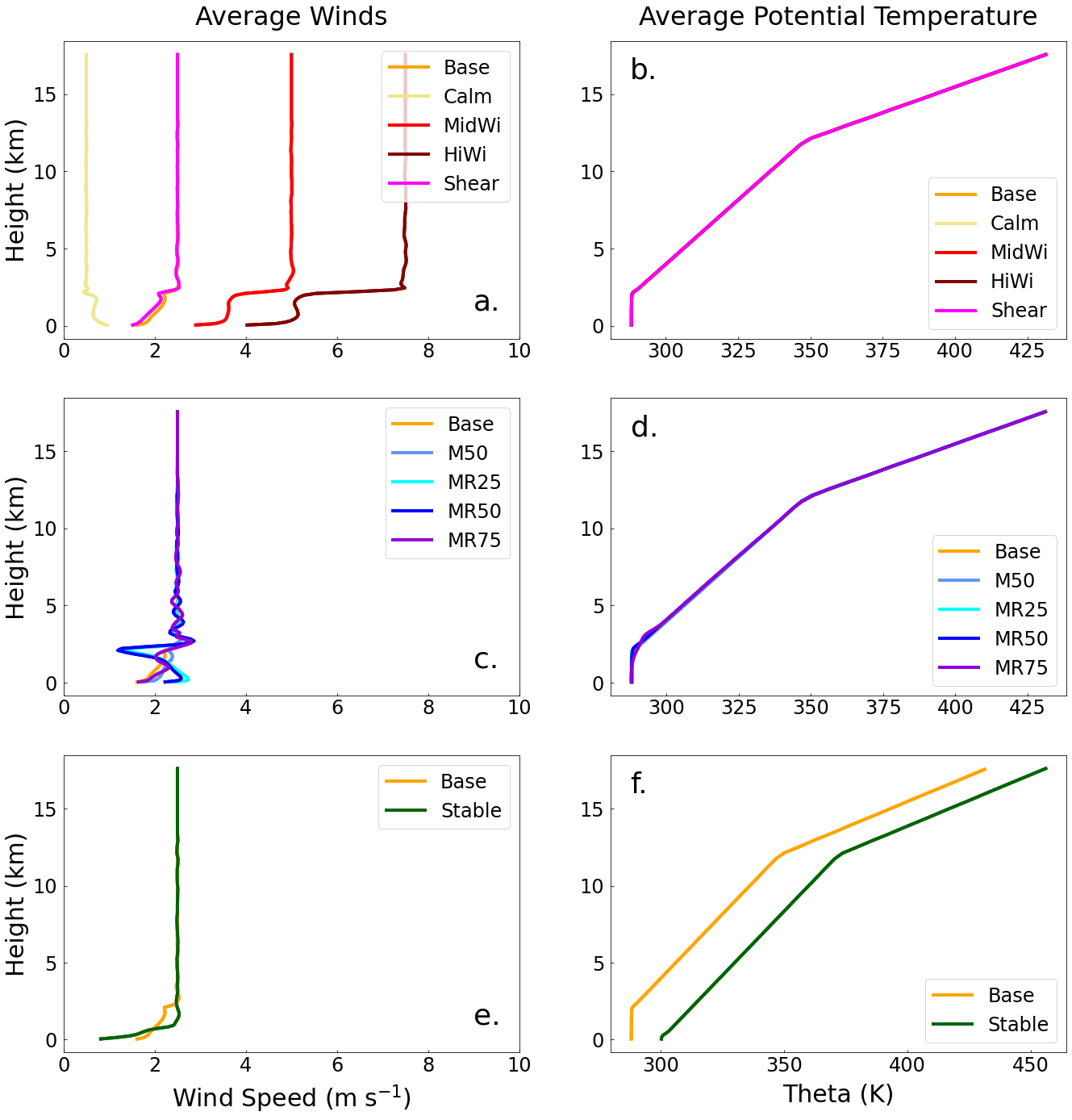}
\caption{Domain-averaged winds in the nest after spin-up for the winds subset (a), moisture subset (c), and stability subset (e). Thermal profiles after spin-up for the winds subset (b), moisture subset (d), and stability subset (f).}
\label{spinup}
\end{figure}

We vary geostrophic wind speed and wind shear, atmospheric moisture, boundary-layer stability, and ignition area in sixteen different simulations to evaluate how meteorological variables influence the smoke distribution generated from two different-sized mass fires. In all cases, the outer domain is initialized and run (spun up) for four hours to achieve equilibrium from the initial condition, after which the interior domain is initialized and run for either 4 or 6 hours, depending on turbulence generation in the nested domain, before the fire is lit. The cell perturbation method \cite{munoz-esparza_generation_2018} is used to generate TKE in the nest, and Figure \ref{spinup} shows the thermal and wind profiles for that domain in each subset of simulations after they have spun up. Once the domains have been spun up, a fire is ignited in the center of the inner domain and burns for 90 minutes. Output from this final segment of the simulations is what we have evaluated in this study.

\begin{figure}[h!]
\centering
\hspace*{-2cm}
\includegraphics[width=11cm]{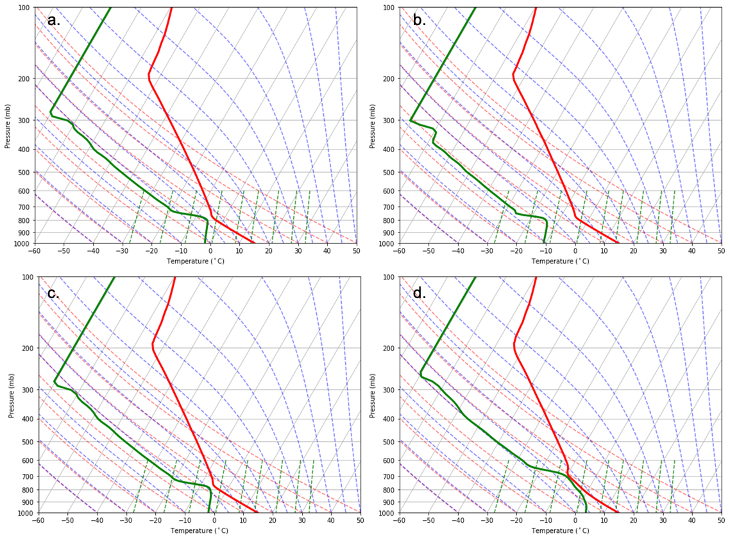}
\caption{Post-spinup soundings for the four moist cases: M50 (a), MR25 (b), MR50 (c), and MR75 (d). Green indicates the dew point temperature profiles, while red indicates the environmental temperature profiles.}
\label{soundings}
\end{figure}

Soundings for the four moist profiles, of which only MR\_75 develops any convective available potential energy (CAPE) after spinup, are presented in Figure \ref{soundings}. CAPE is a measurement of a parcel's potential energy for rise after it reaches the free troposphere, and it is generally used as a metric for the development of deep convection \cite{moncrieff_dynamics_1976,fritsch_numerical_1980}. Before ignition, MR\_75 has an average CAPE value of 41.26 J kg\textsuperscript{-1}.

The base case for each fire size is a low-wind scenario with constant 2.5 m s\textsuperscript{-1} geostrophic winds throughout the domain, no radiation, a dry atmosphere, and a neutrally stratified, 2-km deep boundary layer capped by a stable layer extending through the troposphere (Fig. \ref{spinup}b,d,f). The other cases are divided into subsets based on variations in boundary-layer stability (Stable, with $\frac{d\theta}{dz} = 6\;K\:km^{-1}$ from the top of the boundary layer to the tropopause), winds (Calm, MidWi, HiWi, Shear), or moisture and radiation (M50, MR25, MR50, MR75). Specifications for each simulation can be found in Table \ref{simoverview}. All background winds are applied as geostrophic forcings acting from the ground to the top of the atmosphere. All cases have the same stable thermal profiles above the boundary layer ($\frac{d\theta}{dz} = 6\;K\:km^{-1}$ to the tropopause, and $\frac{d\theta}{dz} = 15\;K\:km^{-1}$ above that). A small amount of fuel moisture (defined as 8\% of the fuel content) is released into the atmosphere upon ignition in all cases.

\begin{table}[ht]
    \centering
    \hspace*{-1.25cm}
    \begin{tabular}{ | M{2cm} | M{1.25cm} | M{1.75cm} | M{1.5cm} | M{1.5cm} | M{1.75cm} | M{1cm} | M{2cm} |} 
        \toprule \toprule \rowcolor{white}
        \textbf{Simulation} & \textbf{ Label} &  \textbf{Wind Speed} &  \textbf{Stability} &  \textbf{Moisture} &  \textbf{Radiation} &  \textbf{Wind Shear} &  \textbf{Relative Humidity} \\ \bottomrule \bottomrule
        Base & Base & 2.5 m s\textsuperscript{-1} & Neutral & Dry & No & No & 0\%\\ \hline
        Calm & CalmWi & 0.5 m s\textsuperscript{-1} & Neutral & Dry & No & No & 0\%\\ \hline
        Mid Winds & MidWi & 5 m s\textsuperscript{-1} & Neutral & Dry & No & No & 0\%\\ \hline
        High Winds & HiWi & 7.5 m s\textsuperscript{-1} & Neutral & Dry & No & No & 0\%\\ \hline
        50\% RH & M50 & 2.5 m s\textsuperscript{-1} & Neutral & Moist & No & No & 50\%\\ \hline
        25\% RH \newline + Radiation& MR25 & 2.5 m s\textsuperscript{-1} & Neutral & Moist & Yes & No & 25\%\\ \hline
        50\% RH + Radiation& MR50 & 2.5 m s\textsuperscript{-1} & Neutral & Moist & Yes & No & 50\%\\ \hline
        75\% RH + Radiation& MR75 & 2.5 m s\textsuperscript{-1} & Neutral & Moist & Yes & No & 75\%\\ \hline 
        Wind Shear & Shear & 2.5 m s\textsuperscript{-1} & Neutral & Dry & No & Yes & 0\%\\ \hline
        Stable & Stable & 2.5 m s\textsuperscript{-1} & Stable & Dry & No & No & 0\%\\ \hline
    \end{tabular}
    \caption{Simulation Overviews. When referenced elsewhere, each label is followed by an underscore and the fire radius. For example, the base case for the 2-km radius fire will be referred to as Base\_2. Note: "Moist," as used in describing the simulations, refers specifically to the addition of atmospheric humidity. Once the fire is lit, all simulations see some release of moisture from the fuel source.}
    \label{simoverview}
\end{table}

Following spinup, either a 2-km radius or 4-km radius fire is ignited in the center of the nest, both lit areally. For all cases, the fire burns for the same amount of time\textemdash{}1.5 hours. In the high-wind case, the fire reaches the domain edge at this point, so the other cases are terminated at the same time to facilitate comparison. 

\subsection{Fuel Source}
A nuclear weapons strike would cause an areal ignition. Key military bases, population-dense cities, or other strategic and anthropogenically-developed locations\textemdash{}all having dense fuel loads\textemdash{}would be likely targets. Observations from World War II, during which over 60 fires were started by incendiary bombing, suggest that firestorms, which generate strong enough convection to loft a great amount of smoke into the upper atmosphere, can develop when the fuel loading exceeds 4 g cm\textsuperscript{-2} \cite{glasstone_effects_1977}. For our simulations, we use heavy logging slash as the fuel bed, which has a smaller loading of 1.3 g cm\textsuperscript{-2}. WRF-Fire has been developed as a wildland fire parameterization, so it offers 13 fuel sources as defined by the Anderson 13 Standard fuel categories \cite{anderson}. Heavy logging slash is the densest option available. A previous study simulating the Hamburg firestorm of WWII used a fuel loading of about 2.6 g cm\textsuperscript{-2} \cite{penner}\textemdash{about two times as large as ours}. Therefore, our simulations may most directly represent forest fire pyroconvection, which can occur in situations with wildland fuel loading \cite{luderer_modeling_2006,trentmann_modeling_2006,fromm,peterson_conceptual_2016}. The 1.3 g cm\textsuperscript{-2} of fuel is also similar to the fuel load assumed by \citeA{reisner} of 0.91 g cm\textsuperscript{-2} \shortcite{robock_how_2019}.

WRF-Fire considers the fuel source to be 100\% combustible, with 2\% of it forming PM2.5 smoke, which then is transported through the domain as a passive tracer. To calculate the smoke profile, all smoke that has exited the domain must be accounted for. An estimate of total smoke (in and out of the domain) at each height can be calculated by assuming that all smoke at the edges of the domain will exit by the next output write time. This time interval has been selected independently for each simulation to minimize error between the smoke generated (an aggregate model variable that updates each time step) and the estimated smoke presence. For the final smoke profile, the smoke is assumed to remain fixed in the vertical after it exits the domain. To calculate the error at each vertical level, the percentage difference between the total smoke generated and the total calculated smoke in (and exited from) the domain is multiplied through the values at each height. This difference is less than 5\% for all cases.


\subsection{Model Uncertainty}

WRF-Fire is under ongoing development; therefore, it has some limitations that could lead to either under-estimation or over-estimation of smoke profile concentrations.  As the smoke is a passive tracer, it does not seed cloud development, nor is smoke scavenged out of the atmosphere via rainfall (both factors leading to overestimation of smoke). Oxygen starvation, which would limit fuel burned or prolong the period of burning, is not considered but is discussed in Section \ref{o2starv} (overestimation). Lightning, which could spark new ignitions, is not simulated (leading to an underestimation of burn area and, therefore, smoke). The wildland fuel density used here is lower than that of urban fuels (underestimation), but the fuel bed is being modeled as continuous, without breaks in the source (overestimation). The fuel burns quicker than urban fuels, leading to an initial burst of heat release and initial lofting, followed by a tapering once the ignition area has burned out, which does not mirror what has been modeled in past studies focused on nuclear winter (underestimation). The fire parameterization has been developed for perimeter and point fires, not areal ignitions as is prescribed in this study (unsure of the consequences). Long-range fire spotting, otherwise known as branding and a method of fire propogation \cite{tarifa_flight_1965,koo_firebrands_2010}, is not currently supported by the model (underestimation). Finally, smoke does not interact with radiation to enable self-lofting behavior (underestimation of lofting and residence time) \cite{robock_climatic_2007,mills_2008,yu}. All of these elements of uncertainty must be taken into consideration when examining the results of these simulations.


\section{Results}
The amount of smoke injected into the upper troposphere and lower stratosphere determines the amount of smoke that may ultimately rise above the tropopause, given enough time and radiative heating. Although not considered here, this process can be simulated in climate models, as shown in \citeA{toon_rapidly_2019}, where smoke injected between 3.5 and 7 km below the tropopause ultimately ascended into the stratosphere. To assess the impacts of local meteorological conditions on smoke transport into the upper atmosphere, we quantify the vertical smoke profile (integrated over the x-y plane to yield Mg m\textsuperscript{-1} of altitude). First, however, we present an analysis of potential oxygen starvation in the model. Following that is the smoke profile analysis, and finally we address some metrics\textemdash{}vertical velocity, CAPE, and horizontal wind speeds\textemdash{}that may be used at the time of combustion to ascertain how high the plume could rise. 


\subsection{Oxygen Starvation in the Model} \label{o2starv}

One point of concern in modeling large fires is the overlooked potential for oxygen starvation. For combustion to occur, a fuel source, a heat source, and oxygen must all be present. Once the amount of oxygen in air drops below 16\%, however, fires can no longer be sustained. Models that do not incorporate chemistry interactions, including WRF-Fire, typically do not consider the potential for oxygen depletion. Therefore, particularly intense, simulated fires may not be physically viable. The initial amount of oxygen in a grid cell with an active fire, combined with the convergence of surrounding air into the fire center, must be large enough to support the heat release and burn rate being calculated. 

To check for oxygen starvation, the mass of oxygen initially present in the lowest two grid cells (ground to 200m) is first calculated according to Equation 1, with an assumed simplified air composition of 79\% nitrogen ($F_{N_{2}}$) and 21\% oxygen ($F_{O_{2}}$) by volume. The molar mass of \(O_2\) is 32 g mol\textsuperscript{-1} ($M_{O_2}$) and the molar mass of \(N_2\) is 28 g mol\textsuperscript{-1} ($M_{N_2}$). Air pressure ($P$)) is given in units of Pa. Each element of the equation is given at the grid cell on the fire mesh with the location (\(i,j\)) at time \(t\).

\begin{equation}
\hspace*{-1.5cm}
O_{2,present,ijt} = (\frac{dP_{ijt}}{g})(dx_{ij})(dy_{ij})\frac{1}{(M_{O_2})(F_{O_{2}})+(M_{N_2})(F_{N_{2}})}M_{O_{2}}F_{O_2}
\end{equation}

The combustion of oxygen is a function of heat release rate. Both \citeA{thornton1917xv} and \citeA{huggett1980estimation} found an average heat release rate ($HRR_{avg}$) of 13.1 MJ kgO\textsubscript{2}\textsuperscript{-1} for the burning of organic solids.  This constant is used to determine the amount of oxygen consumed in accordance with the fire's heat release rate. The oxygen depletion over time is calculated using linear interpolation between heat release rates on 1-minute outputs, with a time step (${\Delta}t$) of 0.1 s. The consumption of oxygen per time step is calculated in Equation 2. $HRR_{ij}$, the heat release rate in each ignited grid cell, is given in W m\textsuperscript{-2}. The variables $dx_{ij}$ and $dy_{ij}$ refer to each cell's extent in the x and y directions. 

\begin{equation}
\hspace*{-1.5cm}
O_{2,cons,ijt} = \sum\limits_{n=0}^{({\Delta}t*600 s)-1} \frac{1}{HRR_{avg}}(dx_{ij})(dy_{ij})({\Delta}t)\biggr((HRR_{ij,t+1}-HRR_{ijt})(\frac{n}{{\Delta}t*600 s})+HRR_{ijt}\biggr)
\end{equation}

The oxygen consumed by combustion is subtracted from the oxygen originally present in the cell each time step, resulting in a cumulative depletion of oxygen. Replenishment of air via inflow to the area is not considered in Equation 2; however, oxygen is assumed to be consumed during this time within a 200-m deep layer.

\begin{figure}[h!]
\centering
\hspace*{-1cm}
\includegraphics[width=16cm]{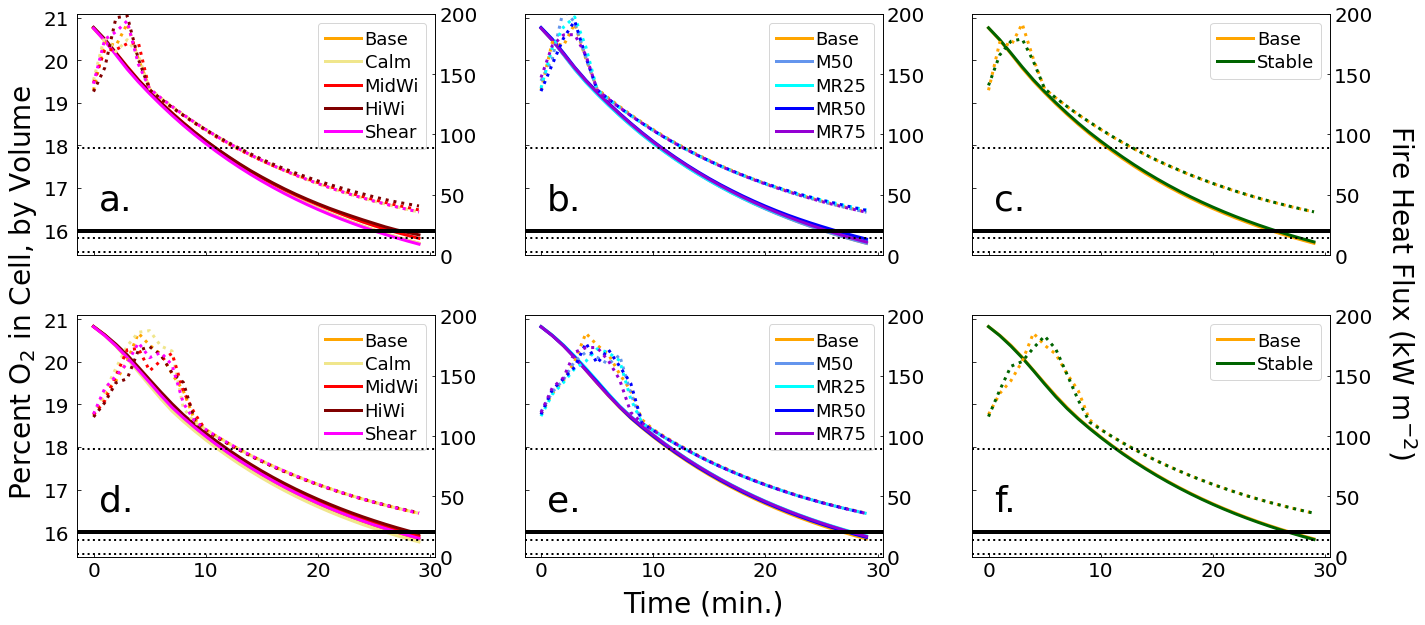}
\caption{\textbf{Top}: Average oxygen depletion in the lowest 200 m of the model across burned area, assuming no air replenishment (left axis, solid), and heat release rate (kW m\textsuperscript{-2}), right axis, dotted) for the 2-km ignition radius fires, grouped by simulation subset (a: Winds, b: Moisture, c: ABL Stability). \textbf{Bottom}: Average oxygen depletion in the lowest 200 m of the model across burned area, assuming no air replenishment (left axis, solid), and heat release rate (kW m\textsuperscript{-2}), right axis, dotted) for the 4-km ignition radius fires, grouped by simulation subset (d: Winds, e: Moisture, f: ABL Stability). The heat release rates for each of \protect\citeA{penner}'s three fire simulations are marked in each plot by horizontal dotted lines (HRR = 2.2 kW m\textsuperscript{-2}, 14 kW m\textsuperscript{-2}, 89 kW m\textsuperscript{-2}). The 16\% depletion threshold is depicted via a thick horizontal line on each plot.}
\label{o2}
\end{figure}

The most intense heat release occurs during the areal ignition of the fires. The fire spreads quickly inward from the outer perimeter, and once the ignition area has been completely burned, what remains is a perimeter fire with a much lower HRR. This process of bulk, rapid fuel consumption takes about four minutes for the 2-km radius fires, and about seven minutes for the 4-km radius fires (Figure \ref{o2}). The greatest oxygen consumption occurs during this time. 

\begin{figure}[h!]
\centering
\hspace*{-1cm}
\includegraphics[width=16cm]{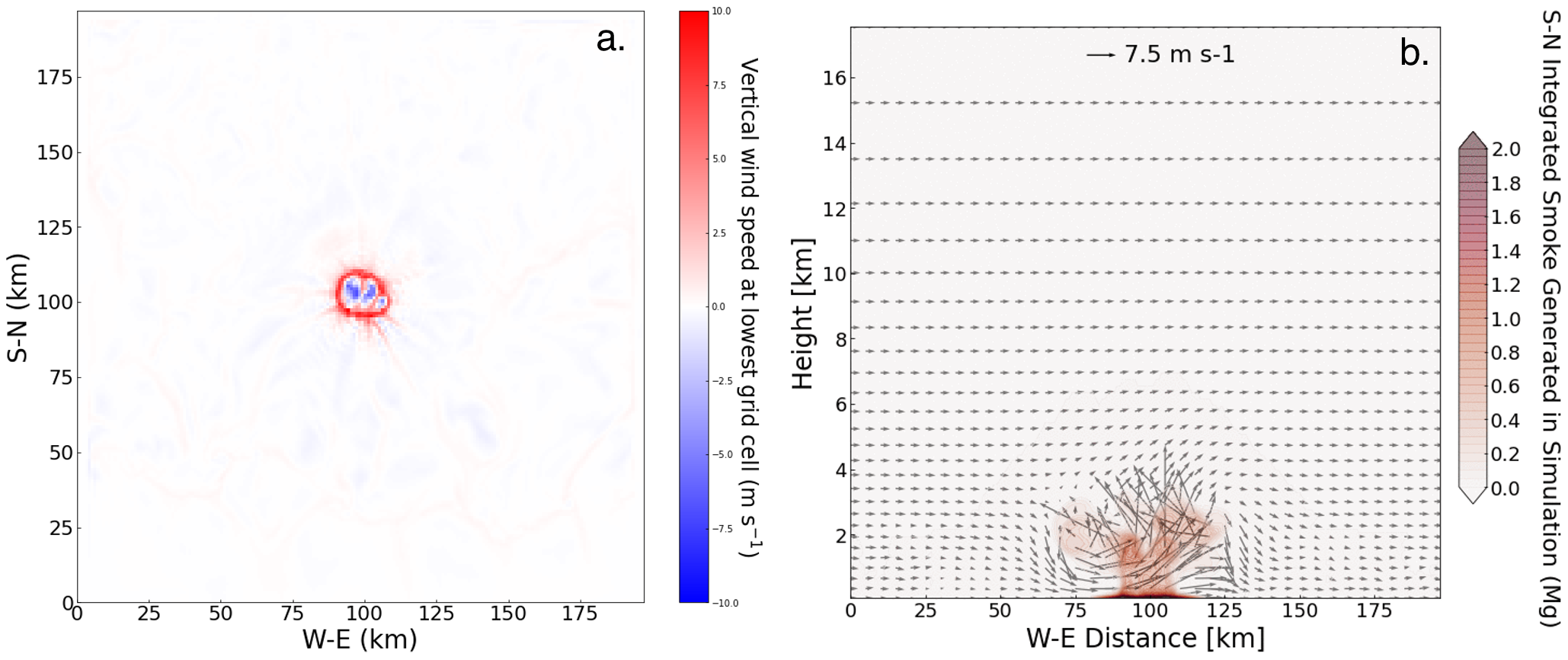}
\caption{Left: Planar view of vertical velocities for Base\_2 at  Right: Average oxygen depletion per grid cell across burned area (solid), assuming no air replenishment, and heat release rate (W m\textsuperscript{-2}, right axis, dotted) for the 4-km ignition radius fires, grouped by simulation subset (Top: Winds, Middle: Moisture, Bottom: ABL Stability).}
\label{o2xsec}
\end{figure}

Assuming a well-mixed boundary layer during and following ignition, we can infer a 30-minute turnover time in the vertical for the lowest 2 km of the model. Therefore, we have only plotted oxygen depletion for this time period in Figure \ref{o2} and assume that turnover following the initial ignition will suffice to maintain burning in what eventually becomes, for all cases, a perimeter fire. O\textsubscript{2} levels drop to just below 16\%, the amount necessary for sustained burning, by minute 30. However, by looking at a cross-sectional plot of winds during this time (Fig. \ref{o2xsec}), we see that O\textsubscript{2} is in fact replenished by horizontal flow into the burn area, at velocities of up to 15 m s\textsuperscript{-1}. With this inflow and considering a full vertical turnover of air in the boundary layer after 30 minutes, oxygen starvation will not be a factor limiting fuel burn. 



\subsection{Smoke Lofting}

Quantification of how meteorology affects the depth of smoke lofting can help better inform assessments of pyrocumulus risk in fire forecasting models and the radiative impacts of large fires in climate models. Therefore, the primary interest of this study is evaluating the sensitivity of smoke lofting to local weather conditions. Here, we examine how varying wind speeds, boundary-layer stability, and moisture affect the immediate rise, or injection height, of smoke into the upper atmosphere. We find that in general, higher wind speeds dampen lofting by dispersing smoke horizontally higher up where the convection column weakens, moisture enhances plume rise due to latent heat release, and a neutral boundary layer provides better thermal conditions for deeper ascension than a stable boundary layer. The quantity of smoke that reaches the upper atmosphere in each simulation is presented in Table \ref{2kmsmk} for the 2-km radius fires and Table \ref{4kmsmk} for the 4-km radius fires. 

\begin{table}[ht]
    \centerline{
    \begin{tabular}{c c c c c c c} 
       \hline
       \multicolumn{1}{|p{2cm}|}{\centering Simulation} & \multicolumn{1}{|p{2cm}|}{\centering Total \\ Smoke (Mg)} & \multicolumn{1}{|p{2cm}|}{\centering Smoke above \\ 9 km (Mg)} & \multicolumn{1}{|p{2cm}|}{\centering Smoke above \\ 9 km (\%)} & \multicolumn{1}{|p{2cm}|}{\centering Smoke above \\ 12 km (Mg)} & \multicolumn{1}{|p{2cm}|}{\centering Smoke above \\ 12 km (\%)} & \multicolumn{1}{|p{2cm}|}{\centering Total Error (\%)} \\ \hline 
       Base\_2 & 3.2E3 & 296.1 & 9.25 & 6.6 & 0.21 & 2.45 \\ 
       CalmWi\_2 & 3.4E3 & 504.1 & 15.03 & 43.3 & 1.29 & -2.26 \\  
       MidWi\_2 & 4.1E3 & 67.2 & 1.64 & 0.5 & 0.01 & -4.27 \\   
       HiWi\_2 & 4.8E3 & 0.6 & 0.01 & 0.0 & 0.00 & -1.13 \\  
       Shear\_2 & 3.2E3 & 279.6 & 8.71 & 6.9 & 0.22 & 2.39 \\   
       M50\_2 & 3.3E3 & 699.4 & 21.17 & 37.3 & 1.13 & 1.20 \\  
       MR25\_2 & 4.0E3 & 287.7 & 7.23 & 4.3 & 0.11 & -3.12 \\ 
       MR50\_2 & 3.5E3 & 518.3 & 14.77 & 33.6 & 0.96 & 2.14  \\
       MR75\_2 & 3.2E3 & 987.9 & 30.79 & 222.6 & 6.94 & 2.21 \\
       Stable\_2 & 3.2E3 & 102.2 &  3.16 & 4.0 & 0.12 & 2.34 \\ 
       \hline
    \end{tabular}}
    \caption{Smoke generation, upper atmosphere and stratospheric smoke lofting, with the 2-km radius ignition area. Negative error values indicate an overestimation of total smoke in the atmosphere, compared with actual smoke generation at the surface, while positive values indicate an underestimation.}
    \label{2kmsmk}
\end{table}

\begin{table}[ht]
    \centerline{
    \begin{tabular}{c c c c c c c} 
       \hline
       \multicolumn{1}{|p{2cm}|}{\centering Simulation} & \multicolumn{1}{|p{2cm}|}{\centering Total \\ Smoke (kg)} & \multicolumn{1}{|p{2cm}|}{\centering Smoke above \\ 9 km (kg)} & \multicolumn{1}{|p{2cm}|}{\centering Smoke above \\ 9 km (\%)} & \multicolumn{1}{|p{2cm}|}{\centering Smoke above \\ 12 km (kg)} & \multicolumn{1}{|p{2cm}|}{\centering Smoke above \\ 12 km (\%)} & \multicolumn{1}{|p{2cm}|}{\centering Total Error (\%)} \\ \hline 
       Base\_4 & 1.3E4 & 394.0 & 3.02 & 103.5 & 0.79 & 0.56 \\ 
       CalmWi\_4 & 1.3E4 & 779.2 & 6.13 & 129.6 & 1.02 & 2.80 \\  
       MidWi\_4 & 1.4E4 & 388.9 & 2.43 & 18.6 & 0.13 & -1.62 \\   
       HiWi\_4 & 1.5E4 & 576.6 & 3.92 & 56.7 & 0.39 & -1.58 \\  
       Shear\_4 & 1.3E4 & 682.7 & 5.09 & 124.2 & 0.93 & -1.71 \\   
       M50\_4 & 1.3E4 & 1921.6 & 14.37 & 134.16 & 1.00 & -1.84 \\   
       MR25\_4 & 1.3E4 & 780.2 & 5.93 & 77.0 & 0.59 & 2.14 \\ 
       MR50\_4 & 1.3E4 & 1972.6 & 14.76 & 235.2 & 1.76 & 0.54 \\
       MR75\_4 & 1.3E4 & 2642.5 & 19.79 & 349.3 & 2.62 & -1.75 \\
       Stable\_4 & 1.3E4 & 251.7 & 1.95 & 99.1 & 0.77 & 1.49 \\ 
       \hline
    \end{tabular}}
    \caption{Smoke generation, upper atmosphere and stratospheric smoke lofting, with the 4-km radius ignition area. Negative error values indicate an overestimation of total smoke in the atmosphere, compared with actual smoke generation at the surface, while positive values indicate an underestimation.}
    \label{4kmsmk}
\end{table}


\subsubsection{Wind Speeds}

Previous studies suggest that wind speeds will affect whether a mass fire develops only as a conflagration, which spreads quickly downwind, or a firestorm, which remains relatively stationary. A firestorm exhibits inward flow of near-surface air from every radial point towards its center; it will generate its own wind systems \cite{countryman,morton}. Firestorms are more likely to consume all the fuel in the ignition area and loft it to higher altitudes than conflagrations. One observational study from WWII suggests that an ambient wind speed of 8 mph, or roughly 3.6 m s\textsuperscript{-1}, is the tipping point between conflagration and firestorm development \cite{rodden}. Slower winds favor the firestorm, while faster speeds favor a conflagration. Due to our lower fuel loading, we do not see development of a firestorm in our Base and Calm cases. We do not see winds flowing into the fires center from all radial directions, even during ignition.

In four simulations for each ignition radius, we vary the geostrophic wind speed, which is kept constant throughout the domain, between 0.5 m s\textsuperscript{-1} and 7.25 m s\textsuperscript{-1}. We also evaluate one slight wind shear case for each ignition size, with speeds increasing from 0 m s\textsuperscript{-1} to 2.5 m s\textsuperscript{-1} in the boundary layer, above which winds are held constant at 2.5 m s\textsuperscript{-1}. There is no specified wind directional shear, except for that which arises due to the Coriolis force (with a Coriolis parameter assumed to be 10\textsuperscript{-4} s\textsuperscript{-1} for a mid-latitude location) and frictional forces.

\begin{figure}[h!]
\centering
\hspace*{-2cm}
\includegraphics[width=16cm]{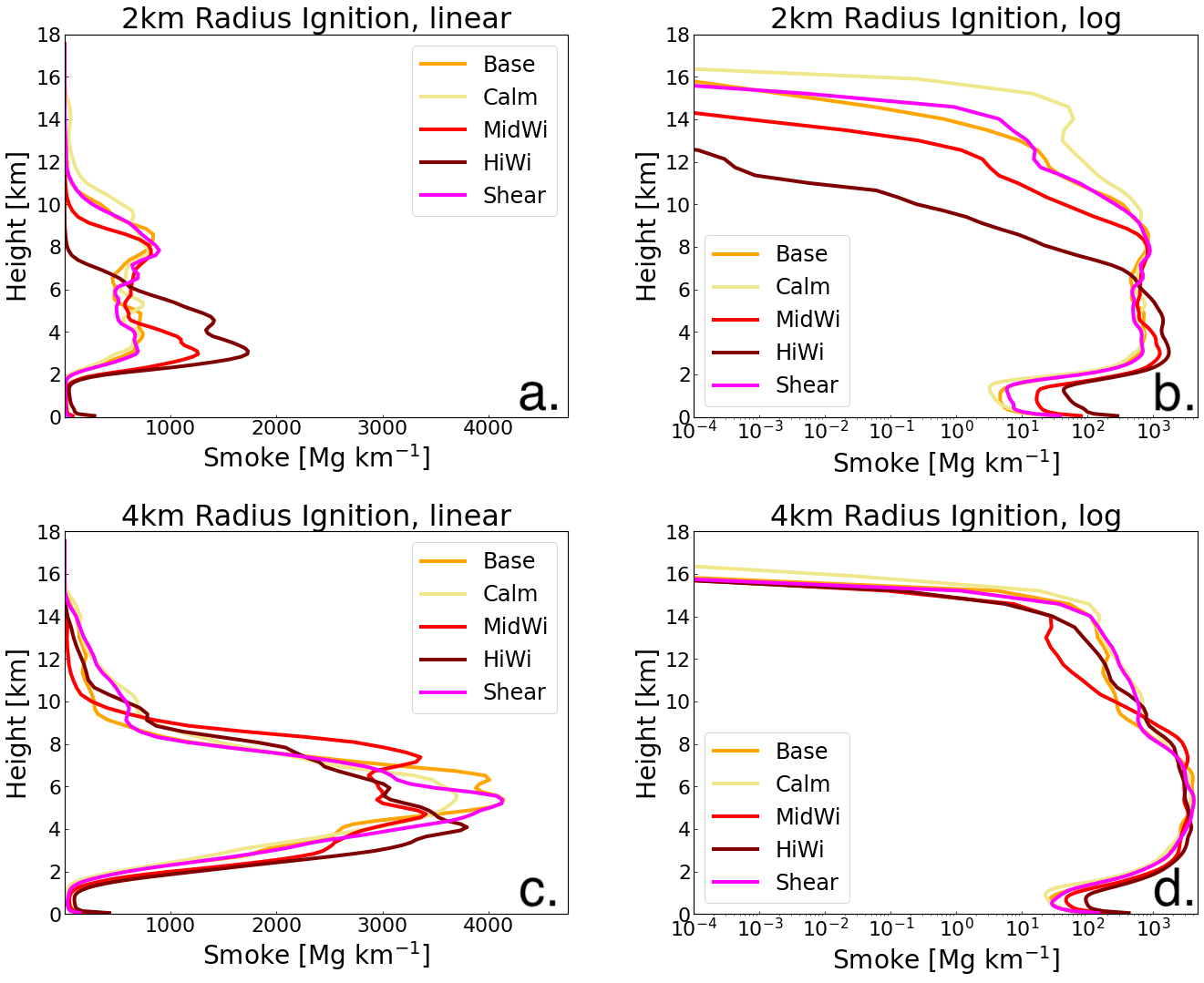}
\caption{Final profiles of vertical smoke distribution (in Mg km\textsuperscript{-1}) for the wind simulations after 1.5 hours of burning. Linear plots (a,c) and logarithmic plots (b,d), for the 2km radius (a,b) and 4km radius (c,d) ignitions are shown.}
\label{4smk_winds}
\end{figure}

\paragraph{2-km Radius Fire}\mbox{}\\\mbox{}\\

\begin{figure}[h!]
\centering
\hspace*{-1cm}
\includegraphics[width=16cm]{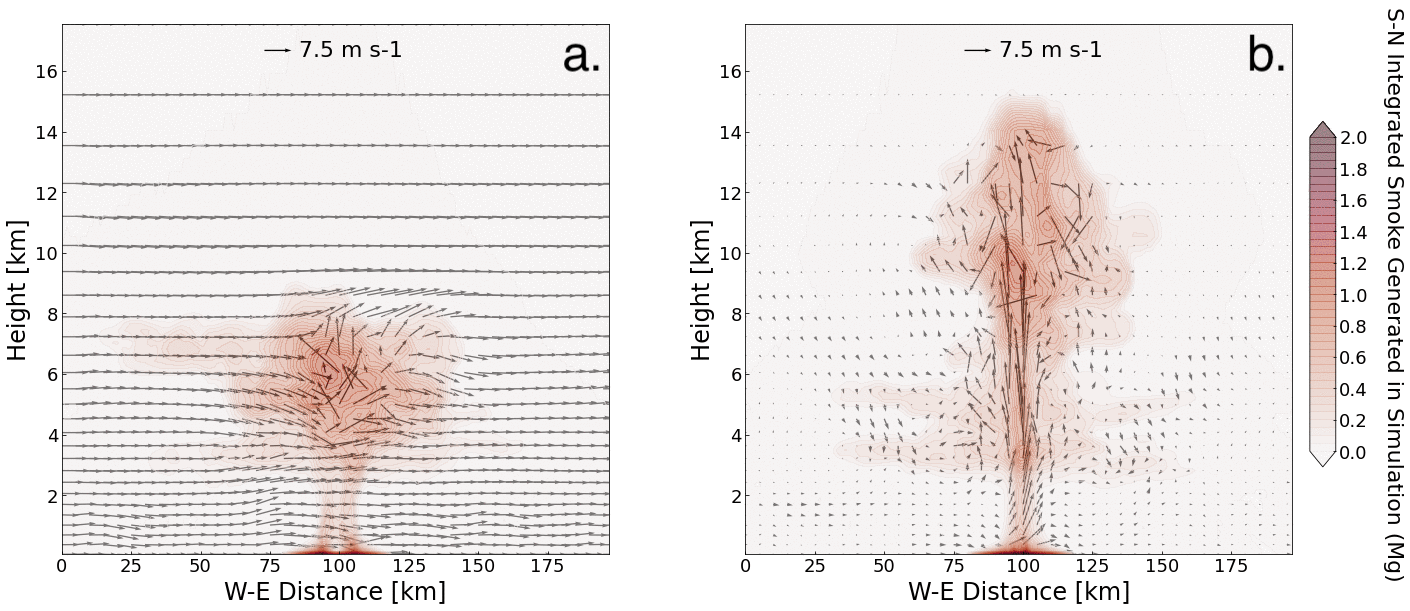}
\caption{Cross-sectional winds taken from a W-E transect through the domain center, with integrated smoke (colored contours), for HiWi\_2 (a) and CalmWi\_2 (b). Stronger horizonal winds in HiWi\_2 dampen the convective lofting of smoke.}
\label{ws_xsec}
\end{figure}

In this set of simulations, the vertical smoke profiles in the 2-km radius fire 1.5 hr after fire initiation demonstrate the impact wind speeds have on lofting (Fig. \ref{4smk_winds}a,b). The largest amount of smoke generation occurs in HiWi\_2, as the faster winds cause the fire to spread more rapidly over the course of 1.5 hours, therefore burning a larger swath of land (5.65\% of the domain, compared with 3.2\% in CalmWi\_2). However, this case also results in a negligible amount of lofting into the stratosphere and the smallest amount of smoke reaching the upper atmosphere, as strong horizontal winds prevent deep vertical motion (Figure \ref{ws_xsec}). Similarly, MidWi\_2 features a nominal amount of smoke reaching beyond tropopause and a small, although not negligible, amount extending above 9 km. Conversely, the weaker wind simulations result in a greater amount of upper atmospheric smoke presence (Fig. \ref{4smk_winds}a,b). In CalmWi\_2, 15\% of the total smoke generated rises above 9 km; Base\_2, with 2.5 m s\textsuperscript{-1} winds, allows 9.25\% of total smoke to rise this high (Table \ref{2kmsmk}). The addition of a small amount of wind speed shear in the boundary layer, as compared with Base\_2, has a slight impact on lofting; smoke above 9 km is reduced, and smoke above the tropopause remains roughly the same.

\paragraph{4-km Radius Fire}\mbox{}\\\mbox{}\\

All cases in the 4-km wind variability simulations result in a greater amount of smoke reaching the upper atmosphere than in the 2-km radius fire, as the increased surface heat flux from the larger ignition area results in stronger convection. As with the 2-km radius fire, the smallest amount of smoke generation and the largest amount of upper atmosphere smoke injection for this subset occurs in the quiescent case (CalmWi\_4). However, compared with the 2-km radius fire, a smaller percentage of total smoke production reaches the upper atmosphere\textemdash{}6.13\%, as compared with 15.03\%. Despite a 400\% increase in the ignition area of the larger fire, the total amount of smoke reaching the upper troposphere and stratosphere increases by 275.1 kg, or only 154\%. Additionally, the percent smoke reaching the stratosphere decreases by 0.2\%, despite an overall increase in total smoke above the tropopause of 86.3 kg, or nearly doubling (Table \ref{4kmsmk}).

The higher wind speed cases\textemdash{}MidWi\_4 and HiWi\_4\textemdash{}both see an increase in percent smoke rise into the upper troposphere and stratosphere with the larger ignition area. HiWi\_4 in particular returns in a greater amount of upper troposphere plume presence when compared with the 2-km radius simulation. Presumably the stronger convection resulting from the larger ignition area provides enough energy to inject more smoke above the inversion at the tropopause. Additionally, the rapid increase in area burned by HiWi\_4 compared with MidWi\_4 contributes to a significantly larger amount of smoke generation, as well as a rise in the fire's heat release rate. As a result, HiWi\_4 has a bigger gain in smoke lofting (Fig. \ref{4smk_winds}). 


\subsubsection{Moisture and Radiation}

The inclusion of moisture and a radiation scheme in the model is important for a realistic simulation. Evaluating the impact of water vapor can provide insight into how significant a role it plays in smoke lofting. The ignition of vegetation releases moisture that can contribute to cloud formation, but the addition of water vapor introduces the potential for pyrocumulonimbus formation, which has been shown to inject smoke into the stratosphere \cite{fromm}.

\begin{figure}[h!]
\centering
\hspace*{-2cm}
\includegraphics[width=16cm]{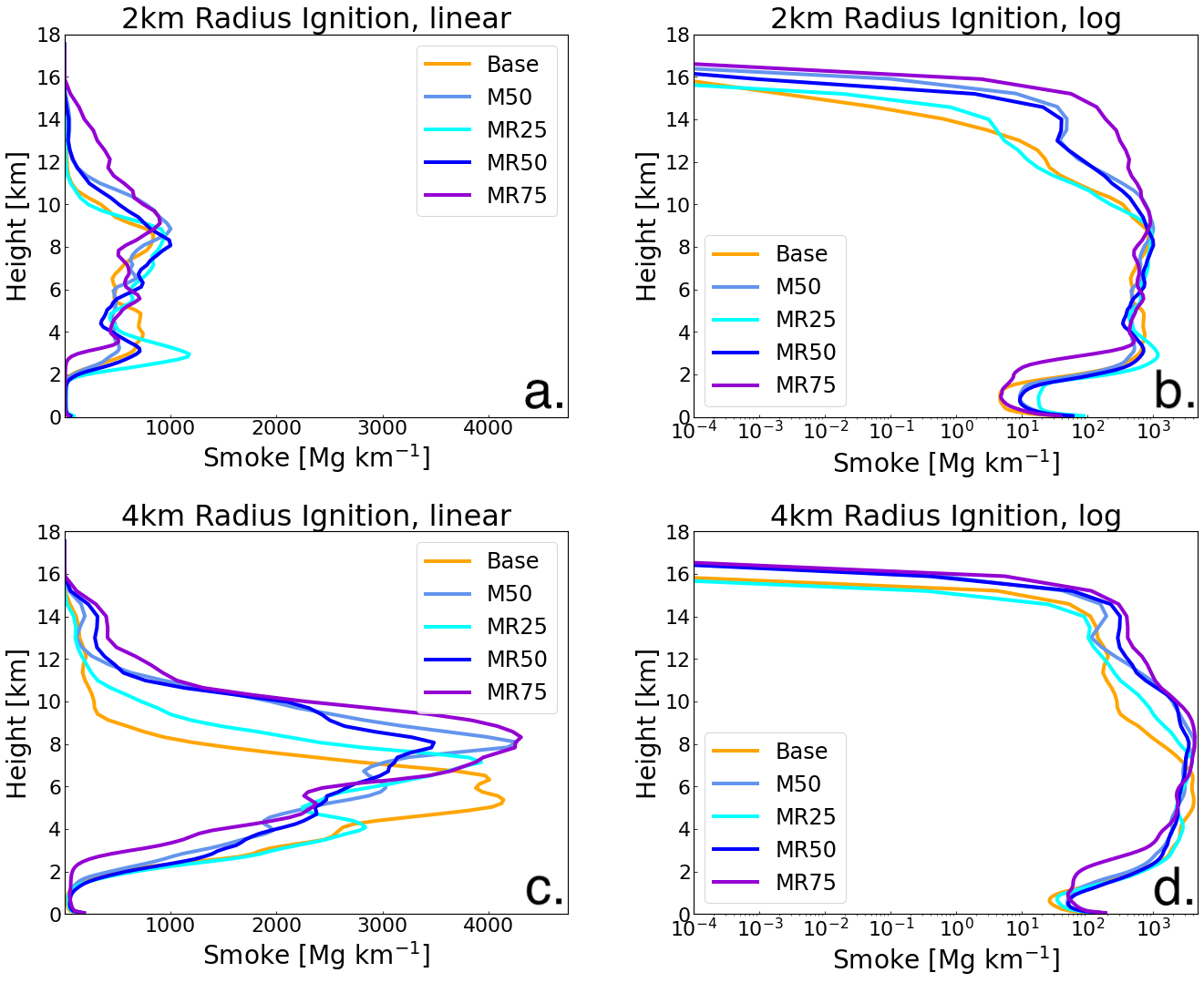}
\caption{Final profiles of vertical smoke distribution (in Mg km\textsuperscript{-1}) for the moist simulations after 1.5 hours of burning. Linear plots (a,c) and logarithmic plots (b,d), for the 2km radius (a,b) and 4km radius (c,d) ignitions are shown.}
\label{4smk_vapor}
\end{figure}

In three simulations for each ignition radius, moisture is varied among a constant 25\%, 50\%, and 75\% relative humidity uniformly throughout the domain, with the RRTMK radiation parameterization enabled. Of note is that the global average humidity in the boundary later is around 75\%. In addition to these simulations, another case with 50\% relative humidity is run without any radiation considered. These simulations are compared alongside the base case, which is dry and does not include radiative effects (Fig. \ref{4smk_vapor}). All simulations have constant 2.5 m s\textsuperscript{-1} winds throughout the domain and a thermal profile mirroring that of the base scenario.

\paragraph{2-km Radius Fire}\mbox{}\\\mbox{}\\

\begin{figure}[h!]
\centering
\hspace*{-1cm}
\includegraphics[width=16cm]{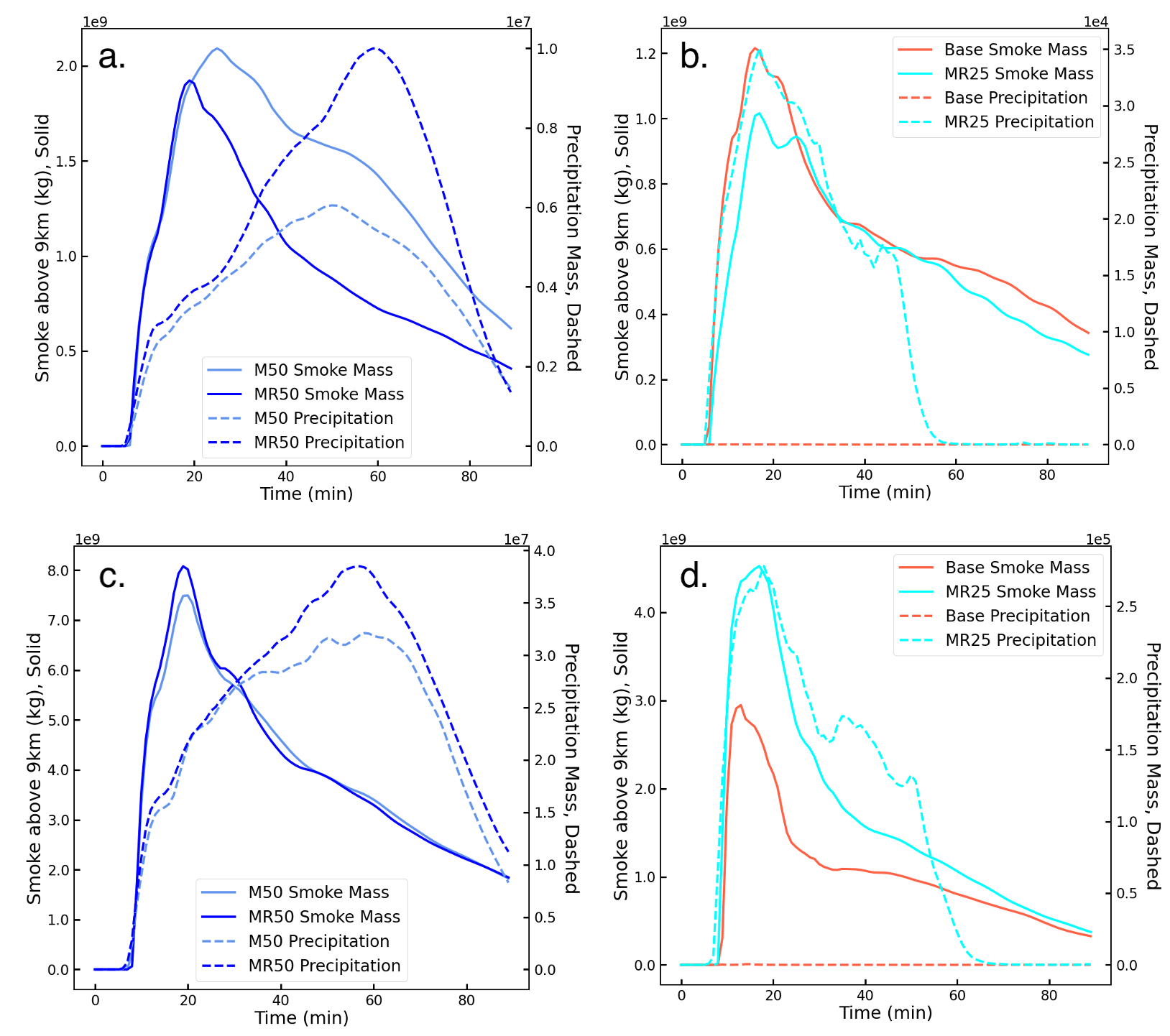}
\caption{Time series of total smoke above 9 km (solid) and precipitation in the domain (dashed) for M50\_2 and MR50\_2 (a), Base\_2 and MR25\_2 (b), M50\_4 and MR50\_4 (c), and Base\_4 and MR25\_4 (d).}
\label{precip}
\end{figure}

In general, adding moisture into the atmosphere enhances air parcel ascent, as that moisture releases latent heat when it rises and condenses. If cloud droplets rise high enough, they freeze into ice\textemdash{a process that also releases heat}. When radiation is considered, a slightly cooler troposphere, as compared with the M50 cases, develops during spinup. Because of this, we see deeper convection for MR50\_2 once the fire is lit, leading to a greater amount of precipitation (here, precipitation is the combination of rain and graupel). Since our passive smoke tracer follows the dynamics of the atmosphere, smoke follows the precipitation out of the upper troposphere, leading to a reduced amount of smoke above 9km and above 12km by the simulations' end (Fig. \ref{precip}a). Similarly, we see a smaller smoke injection above 9km and 12km for MR25\_2 compared with Base\_2. This occurs, again, due to the effects on smoke by precipitation. Base\_2 sees no precipitation development, whereas MR25\_2 does, and as a result a greater amount of smoke exits the upper troposphere and stratosphere in MR25\_2, as compared with Base\_2 (Fig. \ref{precip}b).

M50\_2 lofts over 20\% of its smoke production into the upper troposphere, while MR50\_2 injects only 14.8\% that high. Similarly, Base\_2 results in a 9.25\% plume rise above 9 km, while MR25\_2 only convects 7.2\% to that level. Still, when comparing all the cases using a radiation scheme, an increase in background humidity results in an increase in the amount of smoke reaching the upper troposphere. MR75\_2 sees the greatest amount of smoke lofting of all the 2-km radius ignition simulations, with over 30\% extending above 9 km and 6.9\% rising into the stratosphere (Table \ref{2kmsmk}, Fig. \ref{4smk_vapor}a,b).

\paragraph{4-km Radius Fire}\mbox{}\\\mbox{}\\

When the fire radius is increased to 4 km, the precipitation difference between MR50\_4 and M50\_4 is smaller than that between MR50\_2 and M50\_2 (Fig. \ref{precip}). As a result of this precipitation difference and the deeper convection with radiation, MR50\_4 injects a greater amount of smoke above 9km than MR50\_4 (14.76\%, compared with 14.37\%). Likewise, even though Base\_4 does not develop precipitation while MR25\_4 does, the stronger convection with the larger fire size, deepened by the increased latent heat release in MR25\_4, lofts enough smoke above 9 km that even with rainout more smoke still remains in the upper troposphere in MR25\_4 (5.93\%) than in Base\_4 (3.02\%).

MR75 lofts the most smoke of all the simulations, at almost 20\% above 9 km and 2.6\% above 12 km for MR75\_4, followed by the less humid cases in accordance with their water vapor content (Table \ref{4kmsmk}). Despite significant increases in the total smoke rise, the overall percent plume rise above 9km and 12km for all the moisture cases is lower for the 4-km radius fire than for their counterparts with the 2-km radius fires (Tables \ref{2kmsmk} and \ref{4kmsmk}). 

Following these findings, we may also conclude that even though scavenging effects are not considered in WRF, the effect of scavenging is somewhat represented due to the parcel-following behavior of the passive smoke tracer.


\subsubsection{Atmospheric Boundary Layer (ABL) Stability}

ABL stability, hereafter also referred to as atmospheric stability, may enhance or reduce a large fire's intensity, which in turn affects the probability it will mature into a firestorm \cite{cotton,penner}. Past studies have indicated that a stable atmosphere will dampen convection strength, while a more convective atmosphere will enhance lofting capacity. Two cases \textemdash neutral (the base case) and stable \textemdash have been selected to evaluate this theory. Both have a stable free troposphere, with varying stability in the 2-km deep boundary layer. In the case of both fire sizes, we find that boundary-layer stability exerts a slight influence on the smoke distribution.

\begin{figure}[h!]
\centering
\hspace*{-2cm}
\includegraphics[width=16cm]{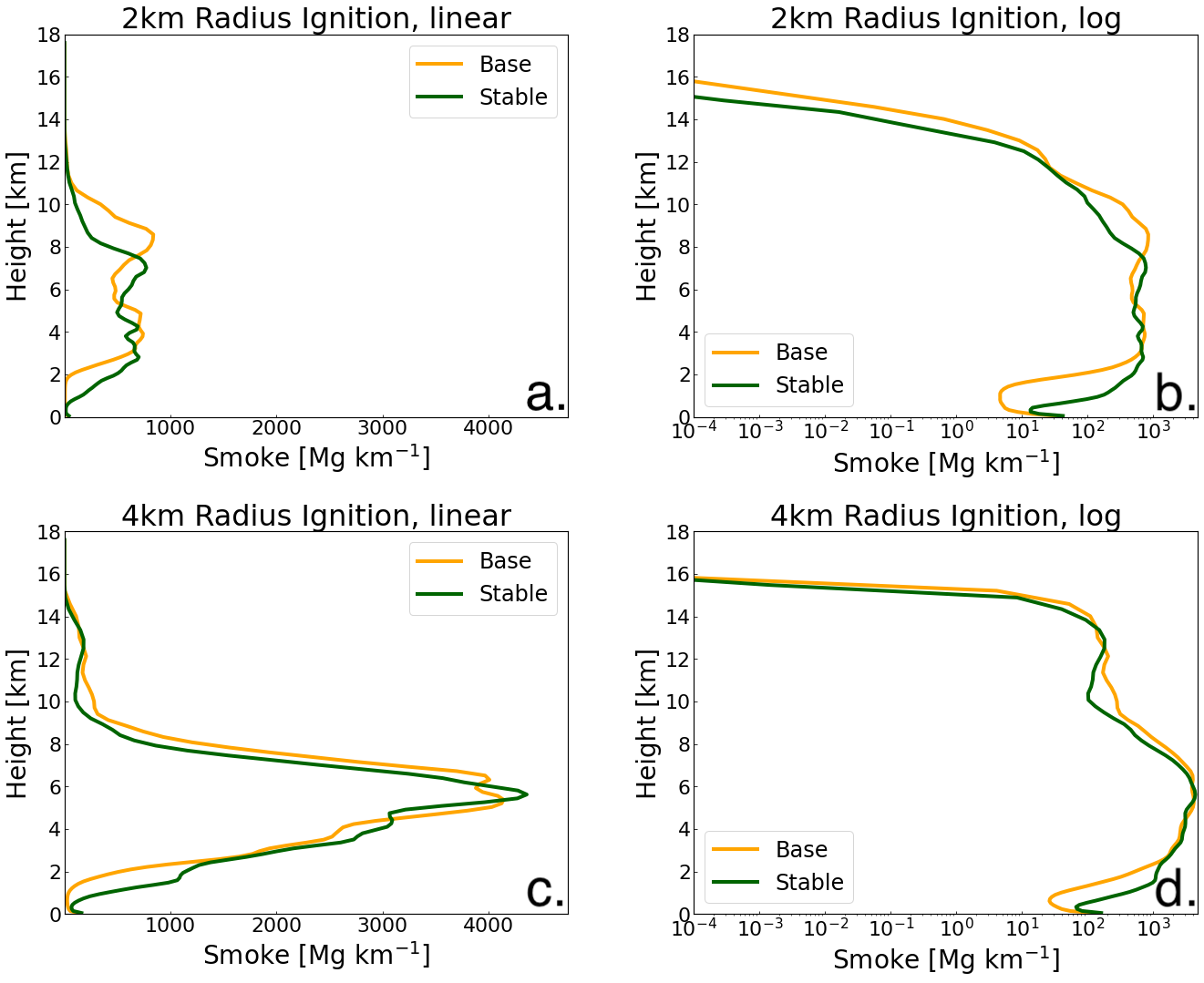}
\caption{Final profiles of vertical smoke distribution (in Mg km\textsuperscript{-1}) for the stability simulations after 1.5 hours of burning. Linear plots (a,c) and logarithmic plots (b,d), for the 2km radius (a,b) and 4km radius (c,d) ignitions are shown.}
\label{4smk_stab}
\end{figure}

With the smaller ignition radius and, therefore, the lower heat release, the stable boundary layer dampens plume rise. Compared with Base\_2, which lofts over 9\% of the smoke generated into the upper troposphere and stratosphere, Stable\_2 is only able to inject 3.16\% that high (Table \ref{2kmsmk}, Fig. \ref{4smk_stab}a,b). However, when the fire radius is increased to 4 km, the difference between the two cases' lofting capabilities shrinks, and Base\_4 lofts only 1.07\% more smoke above 9 km than Stable\_4 (Table \ref{4kmsmk}, Fig. \ref{4smk_stab}c,d). We conclude that, as with the moisture variability, the heat release associated with the larger fire size can overwhelm the influence of atmospheric stability.


\subsection{Relative Importance of Moisture and Winds}

Certain metrics can give us insight into which atmospheric factors more strongly affect the depth of smoke lofting from a fire. Horizontal wind speeds, vertical velocity, and CAPE facilitate comparative analysis among fires in varying atmospheric conditions to evaluate in which settings more smoke may rise high in the atmosphere. 

\subsubsection{Horizontal Wind Speeds and Vertical Velocity}

\begin{figure}[h!]
\centering
\includegraphics[width=12cm]{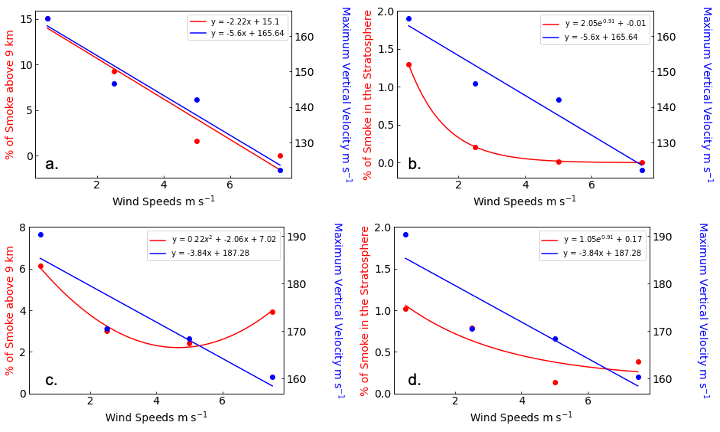}
\caption{Percent of smoke reaching the upper troposphere (a,c) and stratosphere (b,d) compared with wind speeds (red) and maximum vertical velocity in the domain compared with wind speeds (blue), for the 2-km radius ignition (a,b) and the 4-km radius ignition (c,d).}
\label{uv_smkw}
\end{figure}

The simulations for both the 2-km radius and 4-km radius fires have shown that despite fire size, the horizontal wind speeds will influence vertical velocity and, therefore, the total amount of smoke reaching the upper troposphere and stratosphere. With the smaller fire, which has a weaker heat flux and lower values of vertical velocity ($w$), horizontal winds can more easily mix the plume out lower in the atmosphere. The stronger the horizontal winds, the less smoke will be lofted. A clear trend emerges between vertical velocity and horizontal wind speeds (Fig. \ref{uv_smkw}). They have a negative linear relationship with an R-squared value of 0.95. Similarly, vertical velocity and smoke above 9 km also have a linear relationship, with an R-squared value of 0.94 (Fig. \ref{uv_smkw}a). Smoke in the stratosphere and horizontal wind speeds are related via an exponential decay function. 

With an increase in heat flux, the relationship between smoke rise and horizontal winds becomes less clear. Vertical velocity and wind speeds maintain a negative linear relationship, with an R-squared value of 0.84 (Fig. \ref{uv_smkw}c). However, the link between vertical velocity and upper tropospheric and stratospheric lofting, as found here, shifts. They both correlate with horizontal wind speeds via a second-order polynomial, as percent lofting begins to increase with higher wind speeds at this fire size (Fig. \ref{uv_smkw}d). This relationship should be viewed cautiously, as the heat release during ignition slightly differs between the two simulations, as can be seen in Fig. \ref{o2}. The heat flux for HiWi\_4 continually rises, peaks, and then drops, whereas for MidWi\_4, the heat release rate is not a constantly increasing value when the fire is lit. This difference results in a greater amount of initial lofting for HiWi\_4, which is made possible because the convection from this larger fire is less affected by the background horizontal winds. 

Of course, we assess only four data points for each fire size. Therefore, we recommend further simulations incorporating a larger sample size with more ambient horizontal wind speed variability to better quantify these relationships.

\subsubsection{Vertical Velocity and CAPE}
CAPE and vertical velocity ($w$) after ignition can indicate how much smoke is lofted; however, we find that CAPE presents a much stronger signal. CAPE is a measurement of rising potential present in a parcel, based on its buoyancy. Higher CAPE values imply that the parcel will rise. Negative values are associated with stability. Vertical velocity specifies how quickly a parcel is currently ascending or descending. By comparing the maximum values of each, for each simulation, with the percent of smoke lofted into the upper troposphere, a relationship between the metrics and smoke rise emerges. 

\begin{figure}[h!]
\centering
\hspace*{-2cm}
\includegraphics[width=18cm]{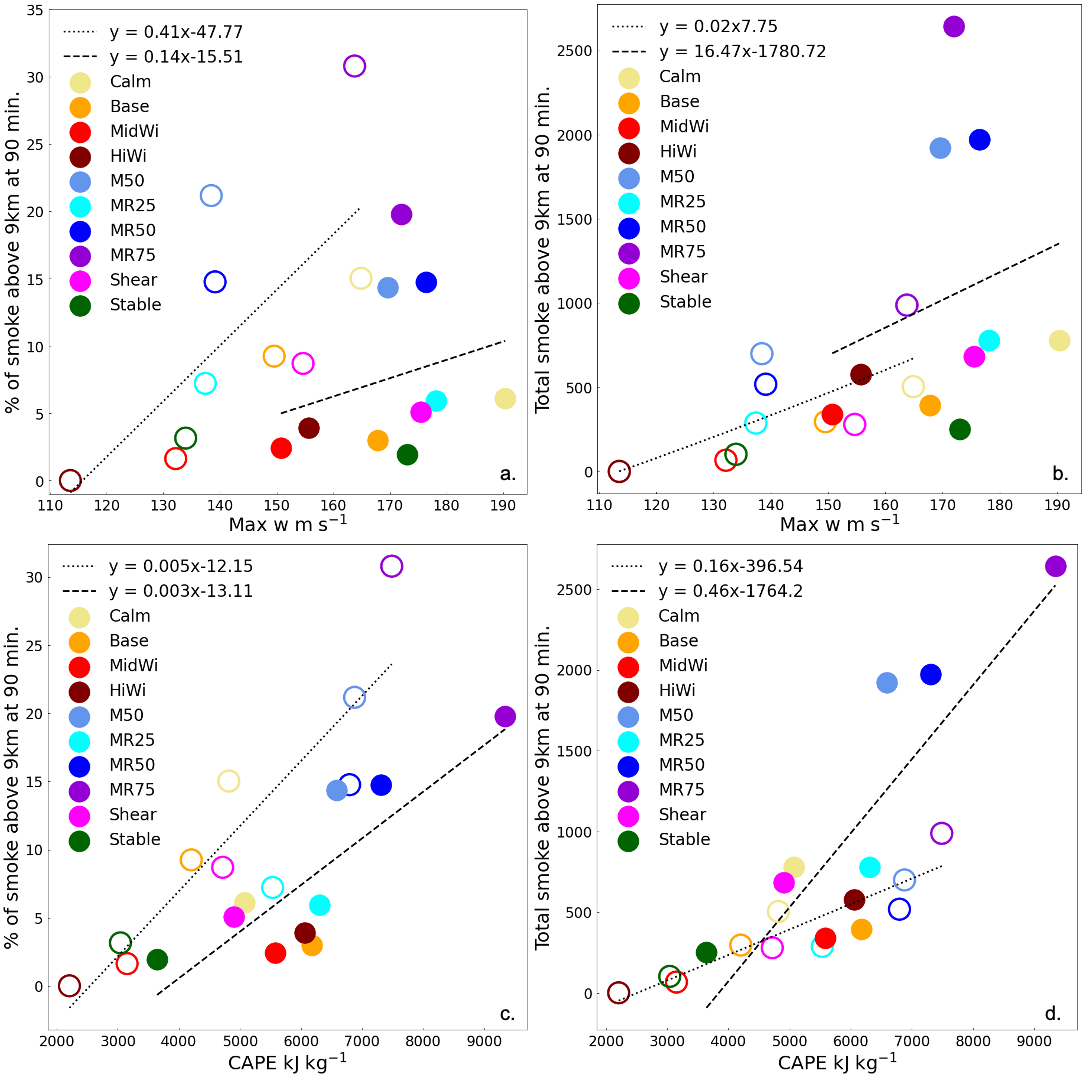}
\caption{Maximum vertical velocity (a,b) and maximum CAPE within the fire area (c,d) compared with \% smoke rise above 9 km (a,c) and total smoke rise above 9 km in Mg (b,d), for all 20 simulations. The unfilled markers indicate values from the 2-km radius ignition, while the solid markers correspond to the 4-km radius fire. Each plot has regressions plotted over their 2-km (dotted) and 4-km (dashed) subsets of data points.}
\label{capew}
\end{figure}

For the smaller fires, as maximum CAPE of any parcel at the surface and maximum $w$ during the simulation increase, the total smoke injected in the upper troposphere generally increases as well (Fig. \ref{capew}). CAPE has a relatively strong correlation with the percent of smoke that reaches the upper troposphere (r\textsuperscript{2} value of 0.8), while vertical velocity has a weaker relationship (r\textsuperscript{2} of 0.46). With the increased fire size and higher HRR, the CAPE-smoke relationship weakens slightly (r\textsuperscript{2} of 0.7), and the $w$-smoke relationship deteriorates altogether.

As is seen in Figure \ref{capew}a and b, the cases with relative humidities at 50\% or greater stand out against what could otherwise be a strong positive relationship between vertical velocity and smoke lofting: higher $w$ would be correlated with increased more smoke rise. The addition of moisture, however, enhances smoke lofting above and beyond the effect of high vertical wind speeds, and the three higher-moisture cases for each fire size have more lofting than what would be indicated by a $w$-smoke correlation. If we consider maximum $w$ as a metric linking smoke lofting with background winds (as that relationship has already been demonstrated), and CAPE as a metric linking lofting with atmospheric humidity, the insensitivity of the CAPE-smoke relationship to varied wind speeds, when compared with the sensitivity of the $w$-smoke relationship to moisture, calls attention to how moisture exerts more influence over plume rise than horizontal winds. 

Additionally, in general, the cases with specified background moisture develop higher CAPE than the dry cases of the same ignition area. However, the dry, quiescent simulations have the highest maximum $w$ values. Although vertical velocities may reach larger values\textemdash{}particularly in situations where horizontal wind speeds are weak\textemdash{}the role of latent heat release and the generation of buoyancy throughout the vertical extent of the domain plays a more dominant role in plume rise than horizontal winds. 

These two findings, therefore, highlight the pivotal role moisture has in dictating plume height. We can conclude that moisture effects are more influential on plume rise than winds.

\section{Discussion}

In this study we have evaluated how local meteorology impacts the depth of smoke injection in the atmosphere as the result of a large, areal fire ignition. Specifically, the effects of varying background wind speeds, relative humidity, and atmospheric stability are examined for two different initial fire areas. Additionally, for the cases with moisture, the effects of a radiation parameterization are also evaluated. Although previous studies \cite{penner} found that plume rise is very sensitive to the fuel burned (and, therefore, the amount of energy released), changes in fuel loading are not considered here. Our simulations all employ a homogeneous wildland fuel bed of about 1.3 g cm\textsuperscript{-2}.

Overall, our results all support the conclusions of most previous studies focused on urban fires following a nuclear conflict. \citeA{rodden}, \citeA{crutzen}, and \citeA{penner} determined that a moist atmosphere and denser fuels enhance lofting, while a stable atmosphere and increased horizontal wind speeds dampen it. In this study, we find that the inclusion of moisture has the most significant impact\textemdash{}more so than either of the other factors assessed\textemdash{}on deep convection, for both fire sizes. A reduction in horizontal wind speeds also enhances plume rise, while the addition of thermal stability in the ABL dampens it.

We have also shown that an increase in fire size does not increase the overall area average heat release rate; it does, however, escalate the overall energy release from the fuel bed upon ignition, as a larger area is consumed by fire. This greater energy release generates overall stronger convection, but it is spread over a wider swath of land. As a result, with the larger fire area, the gains in smoke lofting compared with the increase in smoke generation are smaller, so many of the 4-km simulations actually exhibit a decrease in the percentage of smoke injected into the upper atmosphere.

\begin{figure}[h!]
\centering
\hspace*{-1.25cm}
\includegraphics[width=16cm]{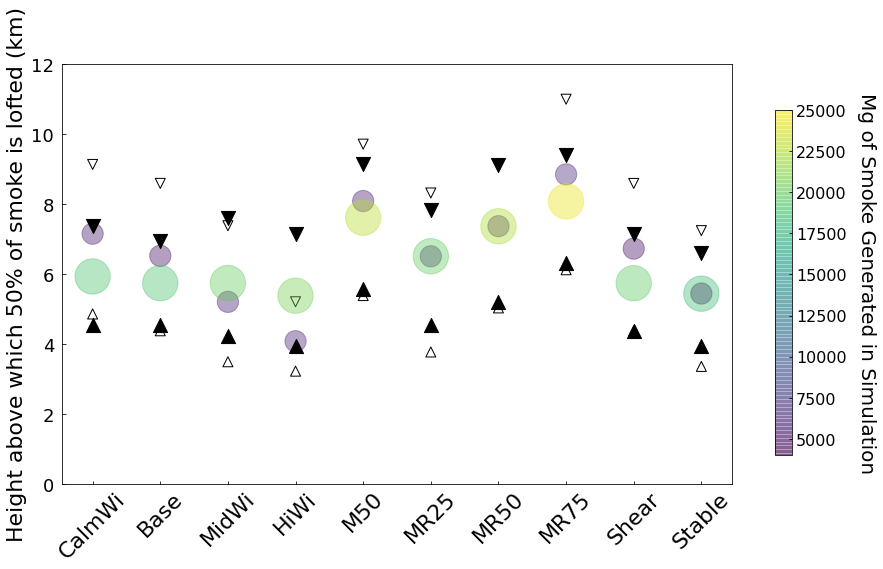}
\caption{Height above which half of the smoke is lofted, based on the final vertical profiles, for each simulation (circles). Smaller markers indicate the 2-km radius fire; larger markers indicate the 4-km radius fire. Markers are colored according to the amount of smoke generated by that simulation. Additionally, the heights above which 25\% and 75\% of the smoke is lofted are indicated by triangles (upward facing for 25\%, downward for 75\%), with smaller, hollow markers signifying the 2-km radius ignition, and larger, filled markers signifying the 4-km radius ignition.}
\label{halfabove}
\end{figure}

Additionally, as convection strengthens (as with the larger fires), factors affecting lofting that are directly impacted by stronger upward motion become less crucial in influencing lofting height. Horizontal wind speeds and boundary-layer stability lose their significance in damping lofting; stronger upward motion allows the dynamics to overcome these elements, and the smoke reaches deeper into the upper atmosphere. And, because the convection column pushes smoke so high and so quickly, the effects of radiation on latent heating are also diminished. Plume injections for these larger fire simulations, therefore, become more comparable with one another\textemdash{}they all push smoke higher up. With respect to variations in water vapor, the increased convection has little impact. The influence a moist atmosphere has on smoke rise is derived from energy release into the convection column as water within a rising parcel condenses. Therefore, increased upward motion does not affect relevance of moisture all that much. These differences (or lack thereof) are depicted in Fig. \ref{halfabove}, where a metric for lofting (the height above which 50\% of smoke is lifted) is plotted for each case and fire size.  

\begin{figure}[h!]
\centering
\hspace*{-1.25cm}
\includegraphics[width=16cm]{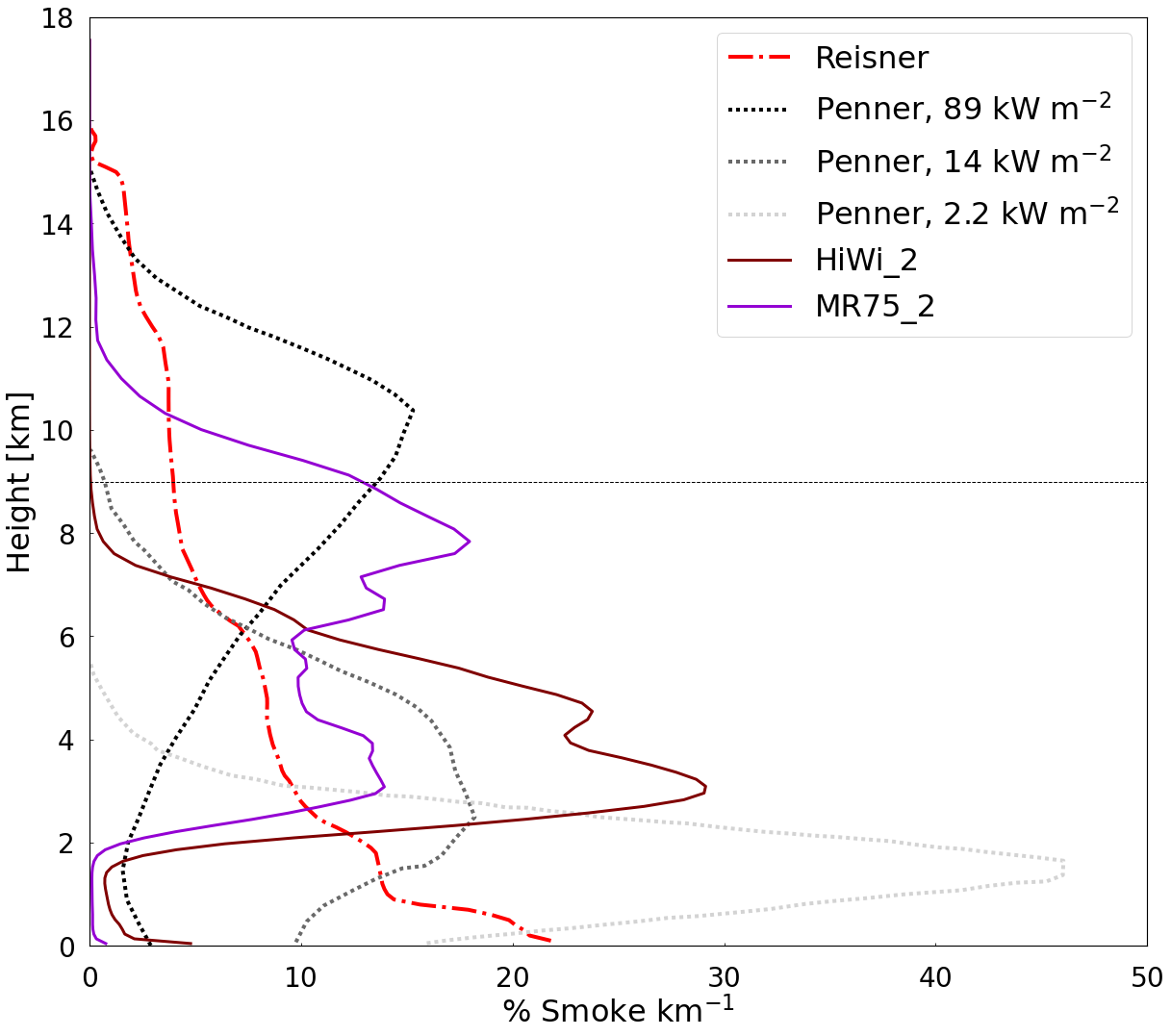}
\caption{Final vertical smoke distributions of fires simulated in \protect\citeA{penner} (dotted), \protect\citeA{reisner} (dash-dotted), and this study (HiWi\_2 and MR75\_2 only, solid). 9 km height is marked by the horizontal dashed line.}
\label{compare}
\end{figure}

We compare our final vertical smoke profiles with those of \citeA{penner} and \citeA{reisner} in Fig. \ref{compare}. Unlike the \citeA{penner} study, we do not have a specified, constant heat release in our fire area, as shown in Fig. \ref{o2}. \citeA{penner} applied heat release rates of 89 kW m\textsuperscript{-2} (10 g cm\textsuperscript{-2} fuel loading), 14 kW m\textsuperscript{-2} (3 g cm\textsuperscript{-2} fuel loading), and 2.2 kW m\textsuperscript{-2} (0.5 g cm\textsuperscript{-2} fuel loading) to simulate three separate fires, with the highest mass loading fire burning for 6 hours, and the other two burning for 12 hours. In contrast, our fires burn for a shorter period of time (1.5 hr) and the surface heat release decreases with time as the fuel in the initial ignition area (mass loading of 1.3 g cm\textsuperscript{-2}) is consumed and the fires begin to spread as perimeter fires. The FIRETEC simulation conducted by \citeA{reisner} is more comparable in that regard. That simulation was run for 40 minutes, and the heat release replicated what would be expected as their fuel depleted with time. 

Interestingly, we find that the distribution of our vertical profiles more closely aligns with those of \citeA{penner} than of \citeA{reisner}. HiWi\_2 injects the majority of its smoke between 3 and 5km, which is akin to \citeA{penner}'s 14 kW m\textsuperscript{-2} case. MR75\_2 injects the majority of its smoke between 6 and 9 km, slightly below \citeA{penner}'s 89 kW m\textsuperscript{-2} injection point that peaks just below 11 km. \citeA{reisner}, on the contrary, saw diminishing smoke injection higher up in the atmosphere, with the bulk of smoke injection occurring within the lowest kilometer of the model. Therefore, compared with the most recent study \cite{reisner}, our results imply a much different risk factor associated with a nuclear conflict. While we did not inform a GCM simulation with the output of our microscale modeling to evaluate the large-scale climate forcing from our fires, we can qualitatively conclude that higher relative humidity and more quiescent conditions at the detonation site, as well as a denser fuel load (associated with a higher heat release rate), could ultimately deliver a higher nuclear winter risk than what was shown by \citeA{reisner}. 

As previously mentioned, the findings in this study are constrained by certain aspects of the model used and the environment selected. Our fuel bed is representative of a wildland source, which has lower fuel density and higher moisture content than urban fuels. The most likely scenario resulting in the ignition of a large, areal fire, as is simulated here, would be the detonation of a powerful weapon, likely in an urban center. WRF-Fire does not currently incorporate cloud seeding, scavenging, or rainout of particulate matter; however, scavenging and rainout are both indirectly captured by the nature of the passive tracer and its parcel-following character. Radiative self-lofting of smoke is not considered. There are no breaks in the fuel bed, nor is there any heterogeneity, as would be found in an urban (or even a mixed wildland) domain. Branding is not considered; lightning is also not modeled. Finally, there is no oxygen starvation within the model, although we demonstrate that oxygen starvation likely does not play a role in our simulations.

This study has examined the influence of several local atmospheric factors on plume rise following the ignition of a large, areal, wildland fire, and has found that background winds, relative humidity, and boundary-layer stability are all important. The motivating context for this idealized sensitivity study is to increase understanding of what local atmospheric factors may prime a region for deep smoke and soot lofting following a nuclear conflict. Therefore, to develop a better understanding of how denser fuel sources may influence smoke lofting, future work with an urban fire model and a global climate model, preferably one considering chemistry and aerosol effects, is recommended.


\acknowledgments
Funding for this study has been provided by the Open Philanthropy Project. This work was authored [in part] by the National Renewable Energy Laboratory, operated by Alliance for Sustainable Energy, LLC, for the U.S. Department of Energy (DOE) under Contract No. DE-AC36-08GO28308. Funding provided by the U.S. Department of Energy Office of Energy Efficiency and Renewable Energy Wind Energy Technologies Office. The views expressed in the article do not necessarily represent the views of the DOE or the U.S. Government. The U.S. Government retains and the publisher, by accepting the article for publication, acknowledges that the U.S. Government retains a nonexclusive, paid-up, irrevocable, worldwide license to publish or reproduce the published form of this work, or allow others to do so, for U.S. Government purposes.

This work utilized resources from the University of Colorado Boulder Research Computing Group, which is supported by the National Science Foundation (awards ACI-1532235 and ACI-1532236), the University of Colorado Boulder, and Colorado State University.


\newpage
\bibliography{agusample}

%
%
%
%
%

\end{document}